\documentclass[english,aps]{revtex4}
\usepackage{graphicx}
\usepackage{amsmath}

\catcode`\@=11
\begin{document}

\title{Two new multi-component BKP hierarchies
\footnote{ Corresponding author:+8610 68916895}\footnote{E-mail
address:wuhongxia@bit.edu.cn}}

\author{Hongxia Wu}
\affiliation{1. Department of Mathematics, Jimei University, Xiamen,
361021, China \\ 2. Department of Mathematics, Beijing Institute of
Technology, Beijing 100081,China }
\author{Xiaojun Liu}
\affiliation{Department of Mathematics, Chinese Agriculture
University, Beijing, PR China}
\author{Yunbo Zeng}
\affiliation{ Department of Mathematical Sciences, Tsinghua
University, Beijing 100084, PR China}

\begin{abstract}
\textbf{Abstract}

We firstly propose two kinds of new multi-component BKP (mcBKP)
hierarchy based on the eigenfunction symmetry reduction and
nonstandard reduction, respectively. The first one contains two
types of BKP equation with self-consistent sources which Lax
representations are presented. The two mcBKP hierarchies both admit
reductions to the $k-$constrained BKP hierarchy and to integrable
(1+1)-dimensional hierarchy with self-consistent sources, which
include two types of SK equation with self-consistent sources and of
bi-directional SK equations with self-consistent sources.
\end{abstract}
\maketitle

\textbf{1.Introduction}

The multi-component KP (mcKP) hierarchy given in [1] contains many
physically relevant nonlinear integrable systems, such as
Davey-Stewartson equation, two-dimensional Toda lattice and
three-wave resonant interaction ones, and attracts a lot of
interests from both physical and mathematical points of view [1-8].
Another kind of multi-component KP equation is the so-called KP
equation with self-consistent sources, which was initiated by V.K.
Mel'nikov [9-11].The first type of KP equation with self-consistent
sources (KPSCS) arises in some physical modes describing the
interaction of long and short wave [8-10,12], and the second type of
KPSCS is presented in [8,11,13]. However, little attention has been
paid to the multi-component BKP hierarchy. Though the first type of
the BKP equation with self-consistent sources (BKPSCS) is
constructed by source generating method [14], the Lax representation
for the first type of BKPSCS and the second type of the BKPSCS have
not been investigated yet.

It is known that the Lax equation of KP hierarchy is given by [15]$$
L_{t_n }  = [B_n ,L] \eqno(1.1)$$where $$ L = \partial  + u_1
\partial ^{ - 1}  + u_2 \partial ^{ - 2}  + \cdots\eqno(1.2)$$is pseudo-differential
operator, $\partial$  denotes $ {\raise0.7ex\hbox{$\partial $}
\!\mathord{\left/
 {\vphantom {\partial  {\partial _x }}}\right.\kern-\nulldelimiterspace}
\!\lower0.7ex\hbox{${\partial _x }$}}$, $ u_i ,\;i = 1,2, \cdots$,
are functions in infinitely many variables $ t = (t_1 ,t_2 ,t_3 ,
\cdots )$ with $ t_1  = x $, and $ B_n  = L_ + ^n $ stands for the
differential part of $L^{n}$.

Owing to the commutativity of $\partial_{t_{n}}$ flows, we obtain
zero-curvature equations of KP hierarchy$$ B_{n,t_k }  - B_{k,t_n }
+ [B_n ,B_k ] = 0\eqno (1.3)$$ Eigenfunction $\Phi$ (adjoint
eigenfunction $\Phi^{\ast}$)satisfy the linear evolution equations
$$\Phi _{t_n }  = B_n (\Phi )\;(\Phi _{t_n }^ *   =  - B_n^ *  (\Phi ^
*  ))\eqno (1.4)$$ The compatibility condition of (1.4) is exactly (1.3).
\\The BKP hierarchy is obtained from the KP hierarchy by imposing the
condition $$L^{\ast}\partial+\partial L=0 \eqno (1.5)$$ The formula
(1.5) implies the vanishing of the even time variables (i.e.,
$t_{2}=t_{4}=\cdots=0 $)and of the constant terms
$B_{n}$,$n=3,5\cdots$, as well as that $u_{2}=-u^{'}_{1}$, $ u_{4} =
- 2u_{3}^{'}  + u_{1}^{(3)}$, $\cdots$, and $\Phi^{\ast}=\Phi^{'}$
for $n$ odd. Taking $k=3, n=5$, (1.3) and (1.5) gives rise to the
BKP equation$$ u_{t _{5} }  + \frac{1} {9}u^{(5)}  - \frac{5}
{9}u_{t_{3} }^{(2)} + \frac{5} {3}uu^{(3)}  + \frac{5} {3}u^{'}
u^{(2)} - \frac{5} {3}uu_{t_{3} }  + 5u^{2} u^{'}  - \frac{5}
{3}u^{'}\partial _{x}^{ - 1} u_{t_{3} }  - \frac{5} {9}\partial
_{x}^{ - 1} u_{t_{3}t_{3} }  = 0 \eqno (1.6)$$ where we use the
notation $u^{(i)}=\frac{\partial^{i}}{\partial x^{i}}$,
$u^{'}=\frac{\partial}{\partial x}$ in this paper.

In this paper, following the idea in [8] and using the eigenfunction
symmetry constraint, we firstly introduce a new type of Lax
equations which consist of the new time $\tau_{k}-$ flow and the
evolutions of wave functions. Under the evolutions of wave
functions, the commutativity of the evolutions of $\tau_{k}-$ flow
and $t_{n}-$ flow gives rise to the first kind of new mcBKP
hierarchy.  This hierarchy enables us to obtain the first and the
second types of BKPSCS and their related Lax representations
directly. This implies that the new mcBKP hierarchy can be regarded
as BKP hierarchy with self-consistent sources (BKPHSCS). Moreover,
this new mcBKP hierarchy can be reduced to two integrable equation
hierarchies: a (1+1)-dimensional soliton equation hierarchy with
self-consistent sources and the $k-$ constrained BKP hierarchy ($k-$
BKPH), which contain the first type and the second type of SK
equation with self-consistent sources and of bi-direction SK
equation with self-consistent sources, respectively. Similar to the
construction of the first kind of mcBKP hierarchy, we can also
construct the second kind of mcBKP hierarchy based on nonstandard
reduction to obtain some new (2+1)-dimensional soliton equation with
self-consistent sources. It is noted that the second kind of mcBKP
hierarchy just as the first kind also admits the $n-$ reduction and
the $k-$ constraint, which lead to some new (1+1)-dimensional
soliton equations with self-consistent sources. Thus, these two
mcBKP hierarchies provide an effective way to find (1+1)-dimensional
and (2+1)-dimensional soliton equations with self-consistent sources
as well as their Lax representations. Our paper is organized as
follows. In section 2, we construct the first kind of new mcBKP
hierarchy based on eigenfuction symmetry constraint and show that it
contains the first and the second types of BKPSCS. In section 3, the
mcBKP hierarchy is reduced to a (1+1)-dimensional soliton hierarchy
with self-consistent source and the $k-$ constrained BKP hierarchy,
respectively. In section 4, the second kind of new mcBKP hierarchy
is also proposed based on nonstandard reduction. In addition, the
$n-$ reduction and the $k-$ constraint of it are also considered. In
section 5, some conclusions are given.

\textbf{2. The first kind of new mcBKP hierarchy}

Following the idea in [8] and using the eigenfunction symmetry
constraint for BKP hierarchy [16], we define $\widetilde{B}_{k}$ by
$$ \widetilde{B}_{k} = B_k  + \sum\limits_{i = 1}^{N} {(r_{i} \partial ^{-1}
q^{'}_{i}  -q_{i} \partial ^{ - 1} r^{'}_{i} )}\eqno (2.1)$$ where
$q_{i},r_{i}$ satisfy (1.4). Then we may introduce a new Lax
equation given by
$$ L_{\tau _{k} }  = [B_k  + \sum\limits_{i = 1}^{N} {(r_{i} \partial ^{-1}
q^{'}_{i}  -q_{i} \partial ^{ - 1} r^{'}_{i} ),L]} \eqno (2.2a)$$
$$ q_{i,t_n } = B_n (q_i ),\;\;r_{i,t_n }  = B_n (r_i ),\;\,i = 1,
\cdots ,N \eqno (2.2b)$$ where $n, k$ are odd. \\\textbf{Lemma 1} $
[B_n ,r\partial ^{ - 1} q^{'} - q\partial ^{ - 1} r^{'}]_ -   =
(r\partial ^{ - 1} q^{'} - q\partial ^{ - 1} r^{'})_{t_n }$
\\\textbf{ Proof:}  Set $B_n =\sum\limits_{i =
1}^{n}a_{i}\partial^{i} $ ($i\geq1$). Then we have $$
\begin{gathered} \ [B_n ,r\partial ^{ - 1} q^{'} - q\partial ^{ - 1} r^{'}]_
-= \sum\limits_{i = 1}^{n} {(a_{i} r^{(i)} \partial ^{ - 1} q^{'}-
a_{i} q^{(i)} \partial ^{ - 1} r^{'}) - \sum\limits_{i = 1}^{n}
{(r\partial ^{ - 1} q^{'} a_{i}\partial ^{i} -q\partial ^{ - 1}
r^{'} a_{i}
\partial ^{i} )_ -  } } \\ = B_{n} (r)\partial ^{ - 1}q^{'} -
B_{n} (q)\partial ^{ - 1}r^{'} - \sum\limits_{i = 1}^{n}(r\partial
^{ - 1} q^{'} a_{i} \partial ^{i}  -q\partial ^{ - 1} r^{'} a_{i}
\partial ^{i} )_ -\end{gathered} $$
\\Applying integration by parts to the second term $$
  \sum\limits_{i = 1}^{n} {(r\partial ^{ - 1} q^{'} a_{i} \partial ^{i}  - q\partial ^{ - 1} r^{'}a_{i} \partial ^{i })_ -  }  =  \cdots  = \sum\limits_{i = 1}^{n} {( - 1)^{i} [r\partial ^{ - 1} (a_{i} q^{'})^{(i)}  - q\partial ^{ - 1} (a_{i}r^{'})^{(i)} ]}
  = r\partial ^{ - 1} B_n^ {*}  (q^{'}) - q\partial ^{ - 1} B_{n}^ {*}
  (r^{'})$$ Noticing the facts that $ q^ *   = q^{'}, r^ *   = r^{'}, q_{t_n }^ *   =  - B_n^ *  (q^ *
  )$ and $r_{t_n }^ *   =  - B_n^ *  (r^ *  ) $, we can complete the
  proof immediately.\\\textbf{Theorem 1. }The commutativity of (1.1) and (2.2a) under (2.2b) leads to the following first kind of new integrable multi-component BKP (mcBKP)
  hierarchy  $$B_{n,\tau _k }  - (B_k  + \sum\limits_{i = 1}^N {(r_i \partial ^{ -
1} q^{'}_i - q_i \partial ^{ - 1} r^{'}_i )} )_{t_n }  + [B_n ,B_k +
\sum\limits_{i = 1}^N {(r_i \partial ^{ - 1} q^{'}_i  - q_i \partial
^{ - 1}r^{'}_i )} ] = 0 \eqno(2.3a)$$ \\ or equivalently $$
\begin{gathered}
  B_{n,\tau _k }  - B_{k,t_n }  + [B_n ,B_k ] + \sum\limits_{i = 1}^N {\{ [B_n ,r_i \partial ^{ - 1} q^{'}_i  - q_i \partial ^{ - 1} r^{'}_i ]} +B_n (q_i )\partial ^{ - 1} r^{'}_i
\\+ q_i \partial ^{ - 1} B^{'}_n (r_i ) - B_n (r_i )\partial ^{ - 1}q^{'}_i
- r_i \partial ^{ - 1} B^{'}_n (q_i )\}  = 0 \end{gathered}\eqno
(2.3a')$$ $$ q_{i,t_n }  = B_n (q_i ),\;r_{i,t_n }  = B_n (r_i
),\quad i = 1, \cdots ,N \eqno(2.3b)$$\\where $n$ and $k$ are
odd.Under (2.3b), the Lax pair for (2.3a) is given by $$ \psi _{t_n
}  = B_n (\psi ),\quad \psi _{\tau _k }  = [B_k  + \sum\limits_{i =
1}^N {(r_i \partial ^{ - 1} q^{'}_i  - q_i \partial ^{ - 1} r^{'}_i
)} ](\psi )\eqno(2.4)$$ \\\textbf{Proof:}We will show that under
(2.3b), (1.1) and (2.2a) lead to (2.3a). For convenience, we assume
$N=1$ and denote $q_{1}, r_{1}$ by $q, r$. By (1.1), (2.2) and lemma
1, we have $$ \begin{gathered}
   B_{n,\tau _k }  = (L_{\tau _k }^n )_ +   = [B_k  + r\partial ^{ - 1} q^{'}-q\partial ^{ - 1} r^{'},\;L^n ]_ +   = [B_k  + r\partial ^{ - 1} q^{'} - q\partial ^{ - 1} r^{'},\;L_ + ^n ]_
   {+}+ [B_k  + r\partial ^{ - 1} q^{'} - q\partial ^{ - 1} r^{'},\;L_ - ^n ]_ +
\\= [B_k  + r\partial ^{ - 1} q^{'}-q\partial ^{ - 1} r^{'},\;L_ + ^n ] -
[B_k  + r\partial ^{ - 1} q^{'} -q\partial ^{ - 1} r^{'},\;L_ + ^n
]_ - +
   [B_k ,L_ - ^n ]_ +  \\ = [B_k  + r\partial ^{ - 1} q^{'} - q\partial ^{ - 1} r^{'},\;B_n ] - [r\partial ^{ - 1} q^{'} -q\partial ^{ - 1} r^{'},B_n ]_ -   + [B_n ,L^k ]_ +
    \\= [B_k  + r\partial ^{ - 1} q^{'} - q\partial ^{ - 1} r^{'},\;B_n ] + (r\partial ^{ - 1} q^{'} -q\partial ^{ - 1} r^{'})_{t_n }  + (B_k )_{t_n }
   \\= [B_k  + r\partial ^{ - 1} q^{'} - q\partial ^{ - 1} r^{'},\;B_n ] + (B_k  + r\partial ^{ - 1} q^{'} - q\partial ^{ - 1} r^{'})_{t_n }
   \end{gathered}$$ \\\textbf{Remark 1.}  (2.3a') and (2.4) indicate that the new mcBKP hierarchy can be regarded as the BKP hierarchy with self-consistent sources and that it is Lax integrable.
\\Next we will list some examples in the first kind of mcBKP
hierarchy.
\\\textbf{Example 1 }(\textbf{The first type of BKPSCS})
For $n=3, k=5$, (2.3)with $u=u_{1}$ leads to the first type of the
BKP equation with self-consistent sources ((2+1)-dimensional CDGKS
equation with self-consistent sources)$$
\begin{gathered}u_{\tau _{5} }  + \frac{1} {9}u^{(5)}  - \frac{5} {9}u_{t_{3} }^{(2)}  +
\frac{5} {3}uu^{(3)}  + \frac{5} {3}u^{'} u^{(2)}  - \frac{5}
{3}uu_{t_{3} }  + 5u^{2} u^{'}  - \frac{5} {3}u^{'} \partial _{x}^{
- 1} u_{t_{3} } - \frac{5}{9}\partial _{x}^{ - 1} u_{t_{3} t_{3} }  + \sum\limits_{i = 1}^{N} {(q_{i}^{(2)} r_{i}  - r_{i}^{(2)} q_{i} )}  = 0, \hfill \\
  q_{i,t_{3} }  = q_{i}^{(3)}  + 3u_{1} q_{i}^{'} ,\;\;r_{i,t_{3} }  = r_{i}^{(3)}  + 3u_{1} r_{i}^{'} ,\quad i = 1, \cdots ,N \end{gathered} \eqno (2.5)$$
\\If we omit the source, we will obtain (2+1)-dimensional CDGKS equation [17].
\\The Lax pair of (2.5) is given by $$ \begin{gathered}
\psi _{t _{3} }  = (\partial ^{3}  + 3u\partial )(\psi ),\\
\psi _{\tau_{5} }  = (\partial ^{5 } + 5u\partial ^{3}  +
\frac{{15}} {2}u^{'} \partial ^{2}  + (\frac{5} {3}\partial _{x}^{ -
1} u_{t _{3} } + \frac{{10}} {3}u^{(2)}  + 5u^{2}  )\partial  +
\sum\limits_{i = 1}^{N} (r_{i} \partial ^{ - 1} q^{'}_{i}  - q_{i}
\partial ^{ - 1} r^{'}_{i} ))) (\psi )
\end{gathered} \eqno(2.6) $$ \\ \textbf{Example 2 (\textbf{The second type of BKPSCS})}
For $n=5, k=3$, (2.3) with $u_{1}=u$ yields the second type of BKP
equation with self-consistent sources $$ \begin{gathered} u_{t_{5} }
+ \frac{1} {9}u^{(5)}  - \frac{5} {9}u_{\tau _{3} }^{(2)}  +
\frac{5} {3}uu^{(3)} + \frac{5} {3}u^{'} u^{(2)}  - \frac{5}
{3}uu_{\tau _{3} } + 5u^{2} u^{'} - \frac{5}
{3}u^{'} \partial _{x}^{ - 1} u_{\tau _{3} }  \hfill \\
   - \frac{5}{9}\partial _{x}^{ - 1} u_{\tau _{3} \tau _{3} }  = \frac{1}
{9}\sum\limits_{i = 1}^{N} {[5q_{i}^{(3)} r_{i}^{'}  - 5r_{i}^{(3)} q_{i}^{'}  + 10q_{i}^{(4)} r_{i}  - 10r_{i}^{(4)} q_{i}} \hfill \\
   + 5(q_{i}^{'} r_{i}  - r_{i}^{'} q_{i} )_{\tau _{3} }  + 30uq_{i}^{(2)} r_{i}  - 30ur_{i}^{(2)} q_{i}  + 30u^{'} q_{i}^{'} r_{i}  - 30u^{'} r_{i}^{'} q_{i} ], \hfill \\
  q_{i,t_{5} }  = q_{i}^{(5)}  + 5uq_{i}^{(3)}  + 5u^{'} q_{i}^{(2)}  + [\frac{5}
{3}\partial _{x}^{ - 1} u_{\tau _{3} }  + \frac{{10}} {3}u^{(2)}  +
5u^{2} + \frac{5}
{3}\sum\limits_{i = 1}^{N} {(q_{i}^{'} r_{i } - q_{i} r_{i}^{'} } )]q_{i}^{'} , \hfill \\
  r_{i,t_{5} }  = r_{i}^{(5)}  + 5ur_{i}^{(3)}  + 5u^{'} r_{i}^{(2)}  + [\frac{5}
{3}\partial _{x}^{ - 1} u_{\tau _{3} }  + \frac{{10}} {3}u^{(2)}  +
5u^{2} + \frac{5}
{3}\sum\limits_{i = 1}^{N} {(q_{i}^{'} r_{i}  - q_{i} r_{i}^{'} } )]r_{i}^{'} \quad  \hfill \\
 \end{gathered} \eqno (2.7)$$ \\ The Lax pair of (2.7) is
given by $$ \begin{gathered}
 \psi _{\tau _{3} }  = [\partial ^{3}  + 3u\partial  + \sum\limits_{i = 1}^{N} {(r_{i} } \partial ^{ - 1} q^{'}_{i}
- q_{i}\partial ^{ - 1} r^{'}_{i} )](\psi ),\\
  \psi _{t_{5} }  =  [\partial ^{5}  + 5u\partial ^{3}  + 5u^{'} \partial ^{2}  + (\frac{5} {3}\partial _{x}^{- 1} u_{\tau
_3 } + \frac{{10}} {3}u^{(2)}  + 5u^{2}  + \frac{{5}}
{3}\sum\limits_{i = 1}^{N}({q^{'}_{i} r_{i}-q_{i}r^{'}_{i} })
)\partial] (\psi )
\end{gathered} \eqno (2.8)$$
\\\textbf{Remark 2} The first type of BKPSCS (2.5) coincide with what obtained in [14] by source generating method. However the Lax representation for (2.5) and the second type of BKPSCS have not been found before.

\textbf{3. The $n-$ reduction and $k-$ constraint of (2.3)}

\textbf{3.1 The $n-$ reduction of (2.3)}
\\The $n-$ reduction of (2.3)is given by [15] $$L^{n}=B_{n}, \ \ or\ \  L^{n}_{-}=0 \eqno
(2.9)$$\\which implies that $$ L_{t_n }  = [B_n ,L] = [L^n ,L] = 0,\
\ B_{k,t_n }  = (L_ + ^k )_{t_n }  = 0,\ \ and \ \ \  q_{i,t_n }  =
r_{i,t_n }  = 0 \eqno (2.10)$$\\If $q_{i}$ and $r_{i}$ are wave
function, they have to satisfy [15] $$ B_n (q_i ) = L^n (q_i ) =
\lambda _i^n q_i ,\;B_n (r_i ) = L^n (r_i ) = \lambda _i^n r_i \eqno
(2.11)$$ \\So it is reasonable to impose the relation (2.11) in the
$n-$ reduction case. By using the Lemma 1 and (2.10), we can
conclude that the constraint (2.9) is invariant under the
$\tau_{k}-$ flow. Due to (2.10) and (2.11), one can drop $t_{n}-$
dependency from (2.3) and get the following (1+1)-dimensional
integrable hierarchy with self-consistent sources $$
\begin{gathered}
  B_{n,\tau _{k }}  + [B_{n} ,B_{k}  + \sum\limits_{i = 1}^{N} {(r_{i} \partial ^{ - 1} q^{'}_{i} - q_{i} \partial ^{ - 1}r^{'}_{i} )} ] =
  0,\\ B_{n }(q_{i} ) = \lambda _{i}^{n} q_{i} ,\ \ B_{n} (r_{i} ) = \lambda _{i}^{n} r_{i} , \ \ i=1,\cdots, N \end{gathered} \eqno
  (2.12)$$\\with the Lax pair given by $$\begin{gathered} B_n (\psi ) = \lambda ^n \psi
  ,\\
   \psi _{\tau _k }  = [B_k  + \sum\limits_{i = 1}^N {(r_i \partial ^{ - 1}q^{'}_i - q_i \partial ^{ - 1}r^{'}_i )} ](\psi ) \end{gathered} \eqno (2.13)$$

\textbf{Example 3 (The first type of SKSCS)} For $n=3, k=5$, (2.12)
present the first type of SK equation with self-consistent sources
(SKSCS) $$ \begin{gathered}
 u_{\tau _{5} }  + \frac{1}
{9}u^{(5)}  + \frac{{5}} {3}u^{'} u^{(2)}  + \frac{5} {3}uu^{(3)} +
5u^{2} u^{'}  +\sum\limits_{i = 1}^{N} {(q_{i}^{(2)} } r_{i} -
q_{i} r_{i}^{(2)} ) = 0, \\
q_{i}^{(3)}  + 3uq_{i}^{'} = \lambda
_{i}^{3} q_{i} \\
r_{i}^{(3)}  + 3ur_{i}^{'}= \lambda
_{i}^{3} q_{i} ,\ \ i = 1, \cdots ,N \end{gathered} \eqno (2.14)$$ \\
(2.13) with $n=3, k=5$ leads to the Lax pair of (2.14)$$
\begin{gathered}(\partial ^{3}  + 3u\partial )(\psi ) = \lambda^{3}
\psi , \\
\psi _{\tau _{5} }  = [\partial ^{5}  + 5u\partial ^{3}  + 5u^{'}
\partial ^{2}  + (\frac{{10}} {3}u^{(2)}  + 5u^{2} )\partial  +
\sum\limits_{i = 1}^{N} {(r_{i}
\partial ^{ - 1} q^{'}_{i}  - q_{i}
\partial ^{ - 1} r^{'}_{i} )} ](\psi ) \end{gathered} \eqno (2.15)$$\\
 \\\textbf{Example 4 (\textbf{The first type
of bi-directional SKSCS)}} For $n=5, k=3$, (2.12) presents the first
type of bi-directional SK equation with self-consistent sources
(bi-directional SKSCS)$$
\begin{gathered} \frac{1}{9}u^{(5)}  - \frac{5} {9}u_{\tau _{3} }^{(2)}  + \frac{5} {3}uu^{(3)}
+ \frac{5} {3}u^{'} u^{(2)}  - \frac{5} {3}uu_{\tau _{3} }  + 5u^{2}
u^{'} - \frac{5}
{3}u^{'} \partial _{x}^{ - 1} u_{\tau _{3} }  \hfill \\
   - \frac{5}{9}\partial _{x}^{ - 1} u_{\tau _{3}\tau _{3} }  = \frac{1}
{9}\sum\limits_{i = 1}^{N} {[5q_{i}^{(3)} r_{i}^{'} }  - 5r_{i}^{(3)} q_{i}^{'}  + 10q_{i}^{(4)} r_{i}  - 10r_{i}^{(4)} q_{i } \hfill \\
   + 5(q_{i}^{'} r_{i}  - r_{i}^{'} q_{i} )_{\tau _{3} }  + 30uq_{i}^{(2)} r_{i}  - 30ur_{i}^{(2)} q_{i}  + 30u^{'} q_{i}^{'} r_{i}  - 30u^{'} r_{i}^{'} q_{i }], \hfill \\
  q_i^{(5)}  + 5uq_{i}^{(3)}  + 5u^{'} q_{i}^{(2)}  + [\frac{5}
{3}\partial _{x}^{ - 1} u_{\tau _{3} }  + \frac{{10}} {3}u^{(2)}  +
5u^{2} + \frac{5}
{3}\sum\limits_{i = 1}^{N} {(q_{i}^{'} r_{i}  - q_{i} r_{i}^{'} } )]q_{i}^{'}  = \lambda _{i}^{5} q_{i} , \hfill \\
  r_{i}^{(5)}  + 5ur_{i}^{(3)}  + 5u^{'} r_{i}^{(2)}  + [\frac{5}
{3}\partial _{x}^{ - 1} u_{\tau _{3} }  + \frac{{10}} {3}u^{(2)}  +
5u^{2} + \frac{5} {3}\sum\limits_{i = 1}^{N} {(q_{i}^{'} r_{i}  -
q_{i} r_{i}^{'} } )]r_{i}^{'}  = \lambda _{i}^{5} r_{i} ,\ \ i = 1,
\cdots ,N \end{gathered}\eqno (2.16)$$ \\ with the Lax pair given by
$$ \begin{gathered} \psi _{\tau _{3} }  = [\partial ^{3}  +
3u\partial  + \sum\limits_{i = 1}^{N} {(r_{i} }
\partial ^{ - 1} q^{'}_{i}  - q_{i}
\partial ^{ - 1} r^{'}_{i} )](\psi ),\\
[\partial ^{5}  + 5u \partial ^{3}  + 5u^{'} \partial ^{2}  +
(\frac{5} {3}\partial _{x}^{ - 1} u_{\tau _{3} } + \frac{10}
{3}u^{(2)}  + 5u^{2}  + \frac{{5}} {3}\sum\limits_{i = 1}^{N}
{(q^{'}_{i} r_{i}-q_{i} r^{'}_{i})}) \partial](\psi ) = \lambda ^{5}
\psi \end{gathered} \eqno (2.17)$$
\\If we take $q_{i}=r_{i}=0$, then (2.14)and (2.16) reduces to the
SK equation and bi-directional SK equation [18].\\\textbf{3.2 The
$k-$ constraint of (2.3)}\\ The $k-$ constraint of (2.3)is given by
[16] $$ L^k  = B_k + \sum\limits_{i = 1}^N {(r_i \partial ^{ - 1}
q^{'}_i - q_i
\partial ^{ - 1} r^{'}_i )} \eqno (2.18)$$ It can seen that (2.18)
together with (2.2) lead to $L_{\tau_{k}}=0$ and $B_{n,\tau_{k}}=0$.
Then (2.3)becomes $k-$ constrained BKP hierarchy $$\begin{gathered}
(B_{k} + \sum\limits_{i = 1}^{N} {(r_{i} \partial ^{ - 1} q^{'}_{i}
- q_{i}\partial ^{ - 1} r^{'}_{i} )} )_{t_{n} }  = [(B_{k}  +
\sum\limits_{i = 1}^{N} {(r_{i}\partial ^{ - 1} q^{'}_{i}  - q_{i}
\partial ^{ - 1} r^{'}_{i} ))_ {+} ^{\frac{n} {k}} } ,B_{k}  +
\sum\limits_{i = 1}^{N} {(r_{i} \partial ^{ - 1} q^{'}_{i}  - q_{i}
\partial ^{ - 1}r^{'}_{i} )} ],\\\ \
q_{i,t_{n} }  = (B_{k } + \sum\limits_{i = 1}^{N} {(r_{i} \partial
^{ - 1} q^{'}_{i} - q_{i} \partial ^{ - 1}r^{'}_{i} ))_ {+}
^{\frac{n} {k}} }(q_{i}) ,r_{i,t_{n }}  = (B_{k}  + \sum\limits_{i =
1}^{N} {(r_{i}
\partial ^{ - 1} q^{'}_{i}  - q_{i} \partial ^{ - 1} r^{'}_{i} ))_ {+} ^{\frac{n}
{k}} }(r_{i}) ,\ \ i = 1, \cdots ,N \end{gathered}\eqno (2.19)$$
\\\textbf{Example 5 (The second type of SKSCS)} For $n=5,k=3$, (2.19) presents the second type of SK equation with self-consistent sources
$$\begin{gathered}
  u_{t_{5} }  + \frac{1}
{9}u^{(5)}  + \frac{5} {3}uu^{(3)}  + \frac{5} {3}u^{'} u^{(2)}  +
5u^{2} u^{'}  = \frac{1}
{9}\sum\limits_{i = 1}^{N} {[5q_{i}^{(3)} r_{i}^{'} }  - 5r_{i}^{(3)} q_{i}^{'}  + 10q_{i}^{(4)} r_{i}  \hfill \\
   - 10r_{i}^{(4)} q_{i}  + 30uq_{i}^{(2)} r_{i}  - 30ur_{i}^{(2)} q_{i}  + 30u^{'} q_{i}^{'} r_{i}  - 30u^{'} r_{i}^{'} q_{i} ], \hfill \\
  q_{i,t_{5} }  = q_{i}^{(5)}  + 5uq_{i}^{(3)}  + 5u^{'} q_{i}^{(2)}  + [\frac{{10}}
{3}u^{(2)}  + 5u^{2}  + \frac{5}
{3}\sum\limits_{i = 1}^{N} {(q_{i}^{'} r_{i}  - q_{i} r_{i}^{'} } )]q_{i}^{'} , \hfill \\
  r_{i,t_{5} }  = r_{i}^{(5)}  + 5ur_{i}^{(3)}  + 5u^{'} r_{i}^{(2)}  + [\frac{{10}}
{3}u^{(2)}  + 5u^{2}  + \frac{5} {3}\sum\limits_{i = 1}^{N
}{(q_{i}^{'} r_{i} - q_{i} r_{i}^{'} } )]r_{i}^{'}, i = 1, \cdots ,N
\end{gathered} \eqno (2.20)$$

\textbf{Example 6 (The second type of bi-directional SKESCS)} For
$n=3, k=5$, (2.19) gives rise to the second type of bi-directional
SK equation with self-consistent sources
$$\begin{gathered} \frac{1} {9}u^{(5)}  - \frac{5} {9}u_{t_{3} }^{(2)}  + \frac{5}
{3}uu^{(3)}  + \frac{5} {3}u^{'} u^{(2)}  - \frac{5} {3}uu_{t_{3} }
+ 5u^{2} u^{'}  - \frac{5}
{3}u^{'} \partial _{x}^{ - 1} u_{t_{3} } , \hfill \\
   - \frac{5}{9}\partial _{x}^{ - 1} u_{t_{3} t_{3} }  + \sum\limits_{i = 1}^{N} {(q_{i}^{(2)} r_{i}  - r_{i}^{(2)} q_{i} )}  = 0, \hfill \\
q_{i,t_{3} }  = q_{i}^{(3)}  + 3u_{1} q_{i}^{'} ,\;\;r_{i,t_{3} }  =
r_{i}^{(3)}  + 3u_{1} r_{i}^{'},\ \ i = 1, \cdots ,N
\end{gathered}\eqno (2.21)$$
Next we will consider the special case of the first type of mcBKP
hierarchy (2.3). Then, starting from the first type one based on the
symmetry eigenfunction constraint, we may obtain one mcBKP hierarchy
based on the nonstandard reduction. Noting that a constant is the
eigenfunction for (1.4), (2.3) with $q_{i}=1$ and $R_{i}=-r^{'}_{i}$
yield the following mcBKP hierarchy $$ \begin{gathered}
  B_{n,\tau _{k} }  - (B_{k}  + \sum\limits_{i = 1}^{N} {\partial ^{ - 1} R_{i} } )_{t_{n} }  + [B_{n} ,B_{k}  + \sum\limits_{i = 1}^{N} {\partial ^{ - 1} R_{i} } ] = 0, \hfill \\
  R_{t_{n} }  =  - B_{n}^ {* } (R) \end{gathered} \eqno (2.22)$$ \\The Lax pair associated with (2.22)
  reads$$ \begin{gathered}
  \psi _{t_{n} }  = B_{n} (\psi ),\ \
  \psi _{\tau _{k} }  = (B_{k } + \sum\limits_{i = 1}^{N} {\partial ^{ - 1} R_{i} } )(\psi )\end{gathered} \eqno
  (2.23)$$ \\where $k$ and $k$ are both odd.

\textbf{  Example 7} For $n=3, k=5$, (2.22)gives rise to $$
\begin{gathered}
  u_{\tau _{5} }  + \frac{1}
{9}u^{(5)}  - \frac{5} {9}u_{t_{3} }^{(2)}  + \frac{5} {3}uu^{(3)} +
\frac{5} {3}u^{'} u^{(2)}  - \frac{5} {3}uu_{t_{3} }  + 5u^{2} u^{'}
- \frac{5} {3}u^{'} \partial _{x}^{ - 1} u_{t_{3} }- \frac{5}
{9}\partial _{x}^{ - 1} u_{t_{3} t_{3} }  + \sum\limits_{i = 1}^{N} {R_{i}^{'} }  = 0, \hfill \\
  \;R_{i,t_{3} }  = R_{i}^{(3)}  + 3(uR_{i} )^{'} ,\quad i = 1, \cdots ,N
  \end{gathered}\eqno (2.24)$$ \\ The Lax pair of (2.24)is given by
$$ \begin{gathered}
  \psi _{\tau _{5} }  = [\partial ^{5}  + 5u\partial ^{3}  + 5u^{'} \partial ^{2}  + (\frac{5}
{3}\partial _{x}^{ - 1} u_{t_{3} }  + \frac{{10}}
{3}u^{(2)}  + 5u^{2} )\partial  + \sum\limits_{i = 1}^{N} {\partial ^{ - 1} R_{i} )} ](\psi ), \hfill \\
  \psi _{t_{3} }  = (\partial ^{3}  + 3u\partial )(\psi
  )\end{gathered} \eqno (2.25)$$ \\\textbf{ Example 8}  For $n=5, k=3$, (2.22)presents $$
 \begin{gathered}
  u_{t_{5} }  + \frac{1}
{9}u^{(5)}  - \frac{5} {9}u_{\tau _{3} }^{(2)}  + \frac{5}
{3}uu^{(3)} + \frac{5} {3}u^{'} u^{(2)}  - \frac{5} {3}uu_{\tau _{3}
} + 5u^{2} u^{'} - \frac{5}
{3}u^{'} \partial _{x}^{ - 1} u_{\tau _{3} }  \hfill \\
   - \frac{5}
{9}\partial _{x}^{ - 1} u_{\tau _{3} \tau _{3} }  = \frac{1}
{9}\sum\limits_{i = 1}^{N }{[\frac{{10}} {9}R_{i}^{(3)}  +
\frac{{10}}
{3}(uR_{i })^{'}  + 5R_{i,\tau _{3} } ]} , \hfill \\
  R_{i,t_{5} }  = R_{i}^{(5)}  + 5(uR_{i}^{(2)} )^{'}  + 5(u^{'} R_{i}^{'} )^{'}  + ([\frac{5}
{3}\partial _{x}^{ - 1} u_{\tau _{3} }  + \frac{{10}} {3}u^{(2)}  +
5u^{2} + \frac{5} {3}\sum\limits_{i = 1}^{N }{R_{i} } ])R_{i} )^{'}
\end{gathered}\eqno (2.26)$$ with the Lax pair given by $$
\begin{gathered}
  \psi _{\tau _{3} }  = [\partial ^{3}  + 3u\partial  + \sum\limits_{i = 1}^{N} {\partial ^{ - 1} R_{i} )} ](\psi ), \hfill \\
  \psi _{t_{5} }  = [\partial ^{5}  + 5u\partial ^{3}  + 5u^{'} \partial ^{2 } + (\frac{5}
{3}\partial _{x}^{ - 1} u_{\tau _{3} }  + \frac{{10}} {3}u^{(2)}  +
5u^{2} + \frac{5} {3}\sum\limits_{i = 1}^{N} {R_{i} } )\partial
](\psi ) \end{gathered} \eqno (2.27)$$\\The $k-$constraint of
(2.22)is given by [19] $$L^{k}=B_{k}+\partial^{-1}R \eqno (2.28)$$
\\ Combining (2.22) with (2.28), we have $$
\begin{gathered}
  (B_{k}  + \partial ^{ - 1} R)_{t_{n} }  = [\{ B_{k}  + \partial ^{ - 1} R\} _ + ^{^{\frac{n}
{k}} } \;,B_{k}  + \partial ^{ - 1} R], \hfill \\
  R_{t_{n} }  =  - [(B_{k}  + \partial ^{ - 1} R)_ + ^{^{\frac{n}
{k}} } ]^ {*}  (R),  i = 1, \cdots ,N \end{gathered} \eqno (2.29)$$
\\\textbf{Example 9} For $n=3, k=5$, we obtain 5-constrained
equation from (2.29)$$ \begin{gathered}
  \frac{1}
{9}u^{(5)}  - \frac{5} {9}u_{t_{3} }^{(2)}  + \frac{5} {3}uu^{(3)} +
\frac{5} {3}u^{'} u^{(2)}  - \frac{5} {3}uu_{t_{3} }  + 5u^{2} u^{'}
- \frac{5} {3}u^{'} \partial _{x}^{ - 1} u_{t_{3} }
   - \frac{5}
{9}\partial _{x}^{ - 1} u_{t_{3} t_{3} }  + \sum\limits_{i = 1}^{N}
{R_{i}^{'} }  = 0,
  \\R_{i,t_{3} }  = R_{i}^{(3)}  + 3(uR_{i} )^{'}\end{gathered} \eqno
  (2.30)$$ \\\textbf{Example 10} For $n=5, k=3$, we obtain 3-constrained
equation from (2.29) $$ \begin{gathered}
  u_{t_{5} }  =  - \frac{1}
{9}u^{(5)}  - \frac{5} {3}uu^{(3)}  - \frac{5} {3}u^{'} u^{(2)}  -
5u^{2} u^{'}  + \frac{{10}} {9}R^{(3)}  + \frac{{10}}
{3}(uR)^{'} , \hfill \\
  R_{t_{5} }  = R^{(5)}  + 5[(uR)^{(3)}  - (u^{'} R)^{(2)}  + (u^{2} R)^{'} ] + \frac{5}
{3}(2u^{(2)} R + R^{2} ) \end{gathered}\eqno (2.31)$$ \\We notice
that (2.31) coincide with what obtained in [19]. \\\textbf{4. The
second kind of new mcBKP hierarchy}

In the previous section, the first kind of new mcBKP hierarchy is
constructed based on the eigenfunction symmetry reduction. In fact,
we find that we can construct the second kind of new mcBKP hierarchy
by using its nonstandard reduction given by [20], which is
completely different from the first kind. \\It is show in [20] that
the non-symmetry constraint $$ L^{k}  = B_{k}  + \sum\limits_{i =
1}^{N} {(r_{i} \partial ^{ - 1} q_{i}^{'}  + q_{i} \partial ^{ - 1}
r_{i}^{'} )},k \ \ is \ \ even \eqno (2.32)$$ \\reduces the BKP
hierarchy to a (1+1)-dimensional integrable hierarchy. So as for the
symmetry constraint, we may define a new Lax equation $$ L_{\tau
_{k} }  = [B_{k}  + \sum\limits_{i = 1}^{N }{(r_{i} \partial ^{ - 1}
q_{i}^{'} + q_{i} \partial ^{ - 1} r_{i}^{'} )} ,\;L] \eqno
(2.33a)$$ $$ q_{i,t_{n} }  = B_{n} (q_{i} ),\;\;r_{i,t_{n} }  =
B_{n} (r_{i} ),\;\,i = 1, \cdots ,N \eqno (2.33b)$$ \\where $n$is
odd and $k$ is even. \\In the exactly same way as for Lemma 1, we
can find \\\textbf{Lemma 2.}$[B_{n} ,r\partial ^{ - 1} q^{'}  +
q\partial ^{ - 1} r^{'} ]_ -   = (r\partial ^{ - 1} q^{'}  +
q\partial ^{ - 1} r^{'} )_{t_{n} }$ \\ \textbf{Theorem 2} (1.1) and
(2.33) lead to the second kind of new integrable mcBKP hierarchy
$$ B_{n,\tau _{k} }  - (B_{k}  + \sum\limits_{i = 1}^{N} {(r_{i} \partial ^{ -
1} q_{i}^{'}  + q_{i} \partial ^{ - 1} r_{i}^{'} )} )_{t_{n} }  +
[B_{n} ,B_{k} + \sum\limits_{i = 1}^{N }{(r_{i} \partial ^{ - 1}
q_{i}^{'} + q_{i}
\partial ^{ - 1} r_{i}^{'} )} ] = 0 \eqno (2.34a)$$ $$
q_{i,t_{n} }  = B_{n} (q_{i}),\;r_{i,t_{n} }  = B_{n} (r_{i} ), i =
1, \cdots ,N \eqno (2.34b)$$ \\ where $n$ is odd and $k$ is even.
With the Lax pair for (2.34a) under (2.34b) given by $$
\begin{gathered}
  \psi _{\tau _{k} }  = (B_{k}  + \sum\limits_{i = 1}^{N} {(r_{i} \partial ^{ - 1} q_{i}^{'}  + q_{i} \partial ^{ - 1} r_{i}^{'} )} )(\psi ), \hfill \\
  \psi _{t_{n} }  = B_{n} (\psi ) \end{gathered} \eqno (2.35) $$
  \\\textbf{Example 11}  For $n=3, k=2$, (2.34) leads to the following new integrable (2+1)-dimensional
  equations $$
  \begin{gathered}
  u_{\tau _{2} }  + u^{(2)}  + \sum\limits_{i = 1}^{N} {(q_{i} r_{i} } )^{(2)}  = 0,\; \hfill \\
   - u_{t_{3} }  + u^{(3)}  + 3uu^{'}  + 3\sum\limits_{i = 1}^{N} {(q_{i}^{'} r_{i}^{(2)}  + r_{i}^{'} q_{i}^{(2)} )}  = 0, \hfill \\
  q_{i,t_{3} }  = q_{i}^{(3)}  + 3uq_{i}^{'} ,\;\;r_{i,t_{3} }  = r_{i}^{(3)}  + 3ur_{i}^{'} ,\quad i = 1, \cdots .N \end{gathered} \eqno
  (2.36)$$ \\ \textbf{Example 12} For $n=3, k=4$,  (2.34) yields to another new integrable (2+1) dimensional
  equation $$
  \begin{gathered}
  3u_{\tau _{4} }  + 2u_{t_{3} }^{'}  + u^{(4)}  + 6(u^{'} )^{2}  + 6uu^{(2)}  + 3\sum\limits_{i = 1}^{N} {(q_{i} r_{i} )^{(2)} }  = 0, \hfill \\
  \frac{2}
{3}u_{t_{3} }^{(2)}  - \frac{4} {3}\partial _{x}^{ - 1} u_{t_{3}
t_{3} } + \frac{2}
{3}u^{(5)}  + 12u^{'} u^{(2)}  + 12u^{'} u^{2}  + 6uu^{(3)}  + 6\sum\limits_{i = 1}^{N} {(q_{i}^{'} r_{i}{(2)}  + r_{i}^{'} q_{i}^{(2)} )}  = 0, \hfill \\
  q_{i,t_{3} }  = q_{i}^{(3)}  + 3uq_{i}^{'} ,\quad r_{i,t_{3} }  = r_{i}^{(3)}  + 3ur_{i}^{'} ,\quad i = 1, \cdots .N \hfill \\
\end{gathered} \eqno (2.37) $$ \\Similar to the first kind of new mcBKP hierarchy, the second kind of mcBKP hierarchy also admits
the $n-$ reduction and the $k-$ constraint given by $B_{n}=L^{n}$
and $ L^{k}  = B_{k}  + \sum\limits_{i = 1}^{N} {(r_{i} \partial ^{
- 1} q_{i}^{'}  + q_{i} \partial ^{ - 1} r_{i}^{'} )}$,
respectively. \\\textbf{Example 13} For $n=3, k=2$, the
2-constrained (2.34) is given by $$ q_{i,t_{3} }  = q_{i}^{(3)}  -
3q_{i} r_{i} q_{i}^{'} , \ \ r_{i,t_{3} }  = r_{i}^{(3)}  - 3q_{i}
r_{i} r_{i}^{'} \eqno(2.38) $$ \\\textbf{Example 14} For $n=3, k=4$,
the 4-constrained (2.34) is given by $$ \begin{gathered}
  u_{t_{3} }  =  - \frac{1}
{2}u^{(3)}  - 3uu^{'}  - \frac{3}
{4}\sum\limits_{i = 1}^{N }{(q_{i} r_{i} )^{'} } , \hfill \\
  q_{i,t_{3} }  = q_{i}^{(3)}  + 3uq_{i}^{'} ,\ \ r_{i,t_{3} }  = r_{i}^{(3)}  + 3ur_{i}^{'} , \ \ i = 1, \cdots .N \hfill \\
\end{gathered} \eqno (2.39) $$ \\ It is noticed that (2.38) is the coupled mKdV equation which has been obtained in [20] and that (2.39) is a new KdV equation with self-consistent sources which is completely different from the two known KdV equations with self-consistent sources introduced by [8,
21-23]. \\ \textbf{Example 15} The 3-reduction of (2.34) for
$n=3,k=2$ reads $$ \begin{gathered}
  u_{\tau _{2} }  - \frac{3}
{2}u^{2}  + \sum\limits_{i = 1}^{N} {(q_{i}^{(2)} r_{i} }  + r_{i}^{(2)} q_{i}  - q_{i}^{'} r_{i}^{'} ) = 0, \hfill \\
  q_{i}^{(3)}  + 3uq_{i}^{'}  = \lambda _{i}^{3} q_{i} ,\;\;\;r_{i}^{(3)}  + 3ur_{i}^{'}  = \lambda _{i}^{3} r_{i} ,\quad i = 1, \cdots .N \hfill \\
\end{gathered} \eqno (2.40)$$ \\\textbf{Example 16} The another 3-reduction of (2.34)
for $n=3,k=4$ reads $$
\begin{gathered}
  3u_{\tau _{4} }  + \frac{1}
{3}u^{(4)}  + 3(u^{'} )^{2}  - 4u^{3}  + 3\sum\limits_{i = 1}^{N} {(q_{i} r_{i}^{(2)} }  + q_{i}^{(2)} r_{i} ) = 0, \hfill \\
  q_{i}^{(3)}  + 3uq_{i}^{'}  = \lambda _{i}^{3} q_{i} ,\;\;\;r_{i}^{(3)}  + 3ur_{i}^{'}  = \lambda _{i}^{3} r_{i} ,\ \ i = 1, \cdots ,N \hfill \\
\end{gathered} \eqno (2.41)$$

\textbf{5. Conclusion}

We firstly propose two new kinds of multi-component BKP hierarchy
based on the eigenfunction symmetry reduction and nonstandard
reduction, respectively. The first kind of mcBKP hierarchy includes
two types of BKP equation with self-consistent sources. It admits
reductions to the $k-$ constrained BKP hierarchy containing the
second type of (1+1)-dimensional integrable soliton equation with
self-consistent sources, and $n-$ reduction of BKP hierarchy
including the first type of (1+1)-dimensional integrable soliton
equation with self-consistent sources. Then we construct the second
kind of mcBKP hierarchy based on nonstandard reduction to obtain
some new integrable (2+1)-dimensional soliton equation with
self-consistent sources. It is noted that the second kind of mcBKP
hierarchy also admits the $n-$ reduction and $k-$ constraint, which
lead to some new integrable (1+1)-dimensional soliton equation with
self-consistent sources. Thus, these two mcBKP hierarchies provide
an effective way to find (1+1)-dimensional and (2+1)-dimensional
soliton equations with self-consistent sources as well as their Lax
representations. Though the solution for the first type of BKPSCS
was constructed by source generating method [14], the solution
structure for the second type of BKPSCS has not been investigated
yet. We will solve the integrable equation in the forthcoming paper.

\textbf{Acknowledgment}

This work is supported by National Basic Research Program of China
(973 Program) (2007CB814800) and National Natural Science Foundation
of China (grant No. 10601028).


\begin{leftline}
\large\bf Reference\normalsize\rm
\end{leftline}

[1]  E Date , M Jimbo, M Kashiwara and T Miwa, J. Phys. Soc. Japan
50 (1981) 3806-3812.

[2] M Jimbo and T Miwa, Publ. Res. Inst. Math. Sci 19 (1983)
943-1001.

[3] Sato M and Sato Y 1982 Soliton equations as dynamical systems on
infinite-dimensional Grassmann

~~~~~manifold. Nonlinear partial differential equations in applied
science (Tokyo).

[4]  Date E , Jimbo M, Kashiwara M and Miwa T 1982 Publ. Res. Inst.
Math. Sci 18 1077-1110.

[5]  V G Kac and J W van de Leur 2003 J. Math. Phys. 44 3245-3293.

[6]Johan van de Leur 1998 J. Math. Phys.39 2833-2847.

[7] Aratyn H, Nissimov E and Pacheva S 1998 Phys. Lett. A 244
245-255.

[8] Liu X J, Zeng Y B and Lin R L 2007 A new multi-component KP
hierarchy (Submitted)

[9]Mel'nikov V K 1983 Lett. Math. Phys. 7 129-136.

[10]Mel'nikov V K 1987Comm. Math. Phys. 112 639-652.

[11] Mel'nikov V K 1988 Phys. Lett. A 128 488-492.

[12] Xiao T and Zeng Y B, 2004 J. Phys. A: Math. Gen. 37 7143-7162.

[13]Wang H Y 2007 Some Studies on soliton equations with
self-consistent sources. PhD thesis,

~~~~Chinese Academy of Sciences.

[14]X B Hu and H Y Wang, Inverse Problem 22 (2006)1903-1920.

[15] L A Dickey, Soliton equation and Hamiltonian systems, (World
Scientific, Singapore, 2003).

[16]I Loris and R Willox, J.Math. Phys., 40 (1999) 1420-1431.

[17] B Konopelchenko and V Dubrovsky, Phys. Lett. A, 102 (1984) 45.

[18] K Sawada and T Kotera, Prog. Theor. Phys., 51(1974) 1355-1367.

[19]Y Cheng, J. Math.Phys. 33 (1992) 3774-3782.

[20]I Loris, J. Phy.A:Math.Gen. 34 (2001) 3447-3459.

[21]V K,Mel'nikov Inverse Problem 6 (1990) 233-246.

[22]J Leon and A Latifi, J. Phys. A: Math. Gen. 23 (1990) 1385-1403.

[23] Y B Zeng, W X Ma and Y J Shao, J. Math. Phy. 42 (2001)
2113-2128.

\end{document}